\documentclass[a4paper,11pt]{article}
\usepackage{pos}
\usepackage{etoolbox}
\usepackage{subcaption}
\usepackage{braket}
\usepackage{appendix}
\usepackage{hyperref}
\usepackage{enumitem}
\usepackage{float}
\usepackage{multirow}
\usepackage{array}

 \DeclareMathOperator{\arccosh}{arccosh} 
 \DeclareMathOperator{\Tr}{Tr}

\begin{document}

\title{
Spectroscopy using tensor renormalization group method
}

\author*[a]{Fathiyya Izzatun Az-zahra}
\author[a]{Shinji Takeda}
\author[b,c]{Takeshi Yamazaki}
\affiliation[a]{Institute for Theoretical Physics, Kanazawa University,\\ Kanazawa 920-1192, Japan}
\affiliation[b]{Institute of Pure and Applied Sciences, University of Tsukuba,\\ Tsukuba 305-8571, Japan}
\affiliation[c]{Center for Computational Sciences, University of Tsukuba,\\ Tsukuba 305-8577, Japan}
\emailAdd{fathiyya@hep.s.kanazawa-u.ac.jp}
\emailAdd{takeda@hep.s.kanazawa-u.ac.jp}
\emailAdd{yamazaki@het.ph.tsukuba.ac.jp}
\abstract{We present a spectroscopy scheme using transfer matrix and tensor network.
With this method, the energy spectrum is obtained from the eigenvalues of the transfer matrix 
which is estimated by coarse grained tensor network of a lattice model,
and the quantum number is classified from the matrix elements of a proper operator 
that can be represented as an impurity tensor network.
Additionally, the momentum of one-particle state and two-particle state 
whose total momentum is zero are classified using matrix elements of proper momentum operators.
Furthermore, using L\"uscher's formula, the scattering phase shift is also computed 
from the energy of two-particle state.
As a demonstration, the method is applied to $(1+1)$d Ising model.}
\FullConference{The 41st International Symposium on Lattice Field Theory\\
The University of Liverpool, United Kingdom\\
28 July - 3 August, 2024 }
\maketitle

\section{Introduction}
Monte Carlo (MC) algorithm has been used for hadron spectroscopy in lattice QCD
and succeeded in reproducing the low-lying hadron spectrum as reported in \cite{snowmass2013}. 
However, this algorithm requires  large time extend to extract the ground state 
and  large statistics to obtain the excited states.
Motivated by these problems,
we look for another computational scheme for spectroscopy 
which is based on tensor network method.

Recently, the research of spectroscopy of a lattice model using tensor network has become progressive.
Some of the research are spectroscopy of (1+1)d quantum electrodynamic (QED) 
with the Hamiltonian tensor network  \cite{Itou:2023img} 
and the computation of ground state energy by Lagrangian tensor network \cite{PhysRevB.107.205123}.
As a continuation of this progress,
we propose a spectroscopy scheme 
using the Lagrangian tensor network combined with transfer matrix.

Unlike the MC method, transfer matrix does not require a large time extend 
and in principle, all of the energy spectrum can be extracted from its eigenvalues.
However, the dimension of this matrix is very large
so that the direct diagonalization is computationally impractical.
To resolve this,  
we use higher order tensor renormalization group algorithm (HOTRG) \cite{PhysRevB.86.045139}
to reduce the dimensionality so that the energy spectrum can be computed but some errors will emerge.
The quantum number and momentum of the energy eigenstate
are identified from matrix elements of a proper operator
that is approximated using HOTRG as well.
Furthermore, the two-particle states with total momentum zero 
which are important to compute the phase shift
are also identified using matrix elements of a chosen operator.
As the first step, 
we carry out the scheme to the (1+1)d Ising model.

\section{Spectroscopy using transfer matrix and tensor network}
In this section, we explain how our scheme can extract the energy spectrum of the (1+1)d Ising model and
identify the quantum number and momentum of the energy eigenstate. 
\subsection{Energy spectrum}
First, we consider the partition function of $(1+1)$d Ising model,
\begin{equation}\label{eq_Z}
Z=\sum_{\{s\}}e^{\beta\sum_{\mathbf{r}\in \Gamma}
\sum_{\mu=1}^2 s(\mathbf{r}+\hat{\mu})s(\mathbf{r})}
\end{equation}
where $\beta=T^{-1}$ is the inverse of temperature, and $s$ is spin with value $s=\pm 1$. 
The spin $s$ resides on two dimensional lattice $\Gamma$ defined as:
\begin{equation}\label{eq_gamma}
\Gamma=\{\mathbf{r}=(t,x)|t=0,1,\ldots, L_t-1
\,\,{\rm and}\,\,
x=0,1,2,\hdots, L_x-1
\}
\end{equation}
where $L_t$ and $L_x$ are system size in time and space direction respectively.
We apply the periodic boundary condition (PBC) to the system so that 
the partition function can be represented as the product of transfer matrix $\mathcal{T}$,
\begin{equation}\label{eq_partitionfunction}
Z=\Tr[\mathcal{T}^{L_t}].
\end{equation}
For $(1+1)$d Ising model this matrix $\mathcal{T}$ is exactly given by 
\begin{equation}\label{eq_tm}
\mathcal{T}_{S'S}=\left(\prod_{x=1}^{L_x-1}\exp[{\beta s(t+1,x)s(t,x)}]\right)
\times
\left(\prod_{x=1}^{L_x-1}\exp\left[{\frac{\beta}{2}s(t+1,x+1)s(t+1,x)}\right]
\exp\left[{\frac{\beta}{2}(t,x+1)s(t,x)}\right]\right).
\end{equation}
The terms in the first parentheses in eq.~(\ref{eq_tm}) is the interaction of spins in the time direction,
while the second one is for the space direction. 
The transfer matrix indices $S'$ and $S$ represent the spin configuration at time $t+1$ and $t$ respectively,
\begin{eqnarray}\label{eq_s's}
S'&=&\{s(t+1,x)|x=0,1,2,\ldots,L_x-1\},\\
S~&=&\{s(t~ ~ ~ ~ ~ ~,x)|x=0,1,2,\ldots,L_x-1\}.
\end{eqnarray}

\begin{figure}[t!]
\centering
\begin{subfigure}[b]{0.4\textwidth}
\centering
\includegraphics[width=4cm,height=4.6cm]{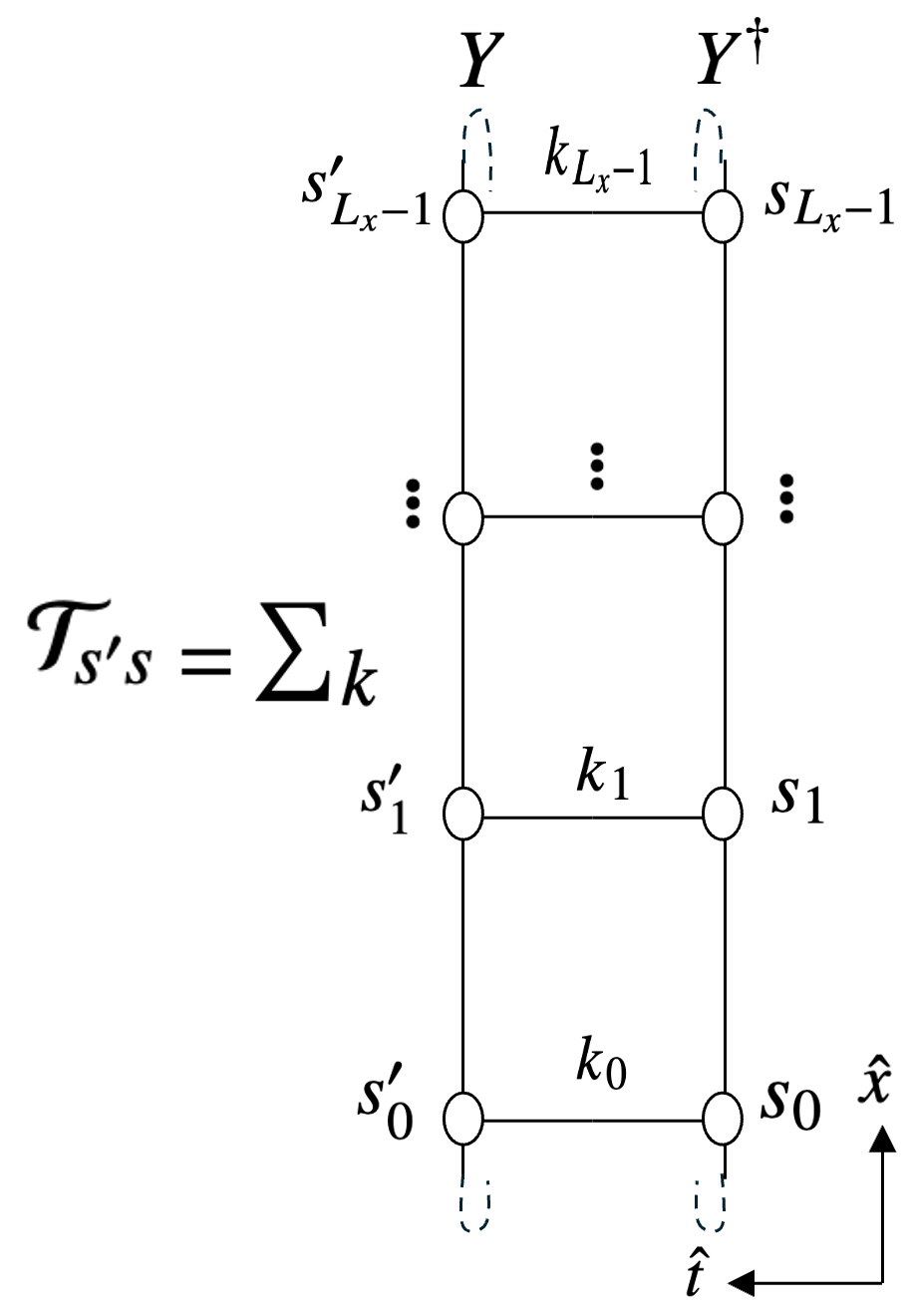}
\caption{}
\label{subfig:T}
\end{subfigure}
\begin{subfigure}[b]{0.4\textwidth}
\centering
\includegraphics[width=7cm,height=4.6cm]{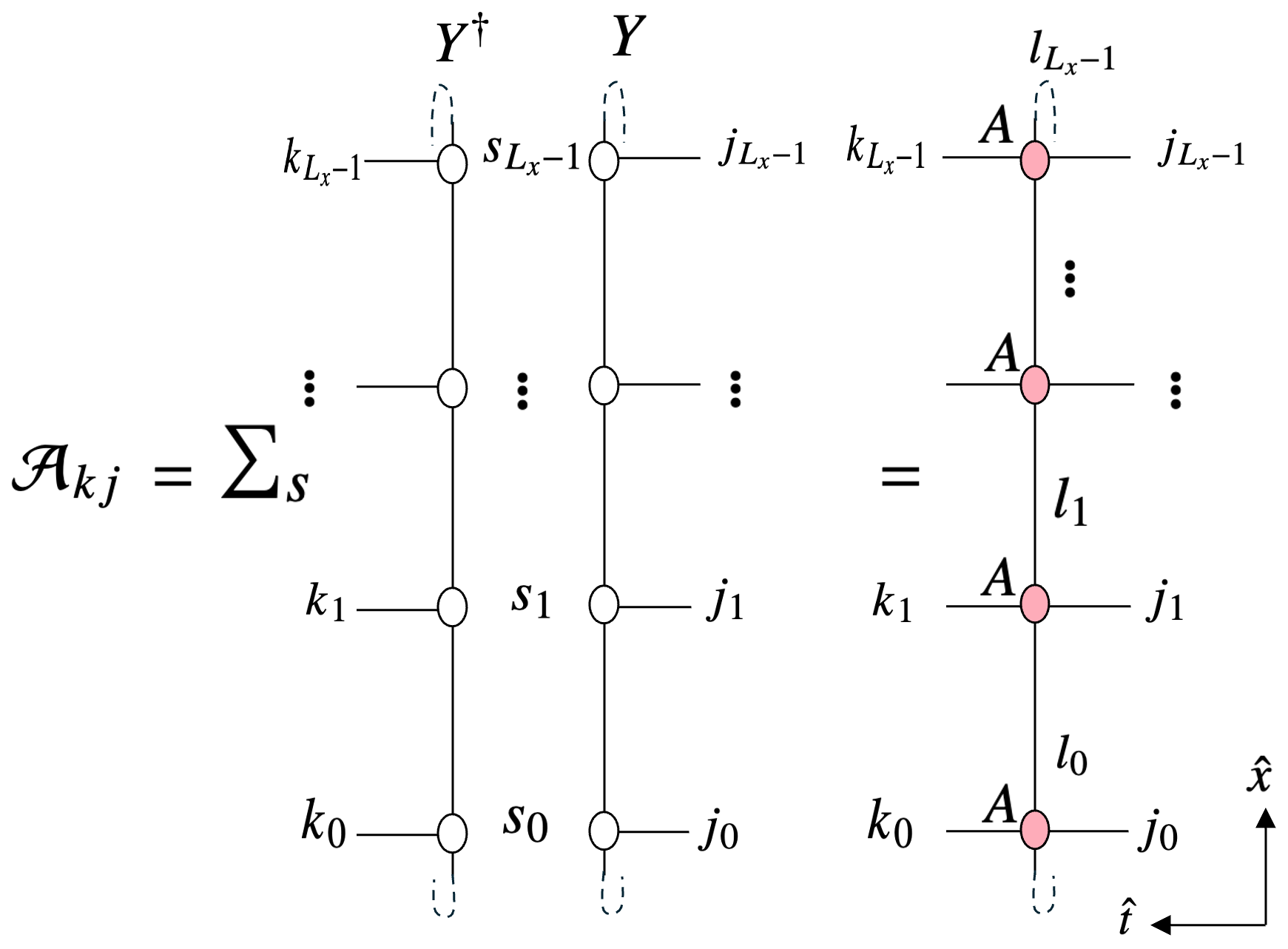}
\caption{}
\label{subfig:A}
\end{subfigure}
\caption{
(a)Transfer matrix $\mathcal{T}$. (b) Tensor network $\mathcal{A}$ 
as a product of matrix $Y$ and $Y^{\dagger}$. 
}
\label{fig:transfer_matrix}
\end{figure}

From the eigenvalue decomposition of the transfer matrix $\mathcal{T}_{S'S}$,
\begin{eqnarray}\label{eq_diag_tm}
T_{S'S}&=&\sum_{a=0}^{2^{L_x}-1}U_{S'a}\lambda_a(U^{\dagger})_{aS},
\end{eqnarray}
we have the unitary matrix $U_{S'a}$ containing eigenvector $|a\rangle$,
and the eigenvalue $\lambda_a$.
The energy of eigenstate $|a\rangle$ denoted as $E_a$ can be extracted 
from $\lambda_a$ as follows:
\begin{equation}\label{eq_lambda}
\lambda_a=e^{-E_a}.
\end{equation}
with $a=0,1,\ldots,2^{L_x}-1$.
Using eq. (\ref{eq_lambda}), 
the energy gap $\omega_a=E_a-E_0$ where $E_0$ is the ground state energy may be written as
\begin{equation}\label{eq_egap2}
\omega_a=\ln \left(\frac{\lambda_0}{\lambda_a}\right)~~~\text{for}~ a=1,2,3,\ldots, 2^{L_x}-1.
\end{equation}
The eq.~(\ref{eq_egap2}) tells us that 
the energy gap $\omega_a$ can be computed directly 
from the eigenvalues of transfer matrix.

Despite of the simpleness, 
the dimension of transfer matrix is equal to the exponential of the system size
so that the direct diagonalization is very hard to be done by computer.
We handle this problem by representing the partition function as a tensor network
since the dimensionality of the tensor network can be reduced into some cut-off $\chi$
using coarse graining process formulated in \cite{PhysRevB.86.045139}. 

Now let us explain how the tensor network can give us the estimation 
of the transfer matrix spectrum $\lambda_a$.
First, we define a matrix $Y$ as follows:
\begin{equation}\label{eq_y}
Y_{S'k}=\prod_{x=0}^{L_x-1}\sum_{k_x=0}^{1}u_{s'_xk_x}\sqrt{\sigma_{k_x}}
\times
\prod_{x=1}^{L_x-1}\exp[{\frac{\beta}{2}s'_{x+1}s'_x}].
\end{equation}
Note that the terms $u$ and $\sigma$ in eq.~(\ref{eq_y}) are obtained from the
singular value decomposition (SVD) of the local Boltzman factor on time direction
\begin{equation} \label{eq_tmy}
\exp[{\beta s'_xs_x}]=\sum_{k_x=0}^{1}u_{s'_xk_x}\sigma_{k_x}u_{k_xs_x}^{\dagger},
\end{equation}
where $s'_x=s(t+1,x)$ and $s_x=s(t,x)$. 
Using the matrix $Y$, the transfer matrix can be rewritten as
 \begin{equation}\label{eq_TM_YYdagger}
 \mathcal{T}=YY^{\dagger}.
 \end{equation}

If we insert eq.~(\ref{eq_tmy}) to eq.~(\ref{eq_partitionfunction}),
then the partition function becomes
\begin{equation}
Z=\Tr\left[\mathcal{T}^{L_t}\right]=\Tr\left[{\left(YY^{\dagger}\right)}^{L_t}\right]
=\Tr\left[\left(Y^{\dagger}Y\right)^{L_t}\right]=\Tr\left[\mathcal{A}^{L_{\tau}}\right]
\end{equation}
where $\mathcal{A}$ is a tensor network defined as
\begin{equation}\label{eq_A_YYdagger}
\mathcal{A}\coloneq  Y^{\dagger}Y.
\end{equation}
The graphical images of eqs.~(\ref{eq_A_YYdagger}) and (\ref{eq_TM_YYdagger}) 
are given in Fig.~\ref{fig:transfer_matrix}.
From this figure, 
we see that the tensor network $\mathcal{A}$ can be written as a product of rank-four tensor $A$ defined as
\begin{equation}
A_{k_xl_xj_xl_{x-1}}\coloneq \sqrt{\sigma_{k_x}\sigma_{l_x}\sigma_{j_x}\sigma_{l_{x-1}}}
\sum_{s_x}(u^{\dagger})_{k_xs_x}(u^{\dagger})_{l_xs_x}
u^{\dagger}_{s_xj_x}u^{\dagger}_{s_xl_{x-1}}.
\end{equation}
The SVD of matrix $Y$ is given by $Y=U\sqrt{\lambda}W^{\dagger}$
where $U$ and $\lambda$ are unitary matrix and eigenvalues of transfer matrix in eq.~(\ref{eq_diag_tm}).
If we subtitute the SVD of $Y$ into eq.~(\ref{eq_TM_YYdagger}), 
then we get eq.~(\ref{eq_diag_tm}).
On the other hand, if the SVD of $Y$ is inserted to eq.~(\ref{eq_A_YYdagger}),
we obtain the EVD of  $\mathcal{A}$:
\begin{equation}\label{eq_evd_A}
 \mathcal{A}=W\lambda W^{\dagger}.
 \end{equation}
Eqs. (\ref{eq_diag_tm}) and (\ref{eq_evd_A}) show that 
the tensor network  $\mathcal{A}$ has the same eigenvalue 
as the transfer matrix $\mathcal{T}$ but with different eigenvectors.

\begin{figure}[t!]
\centering
\begin{subfigure}[b]{0.4\textwidth}
\centering
\includegraphics[width=4cm,height=4cm]{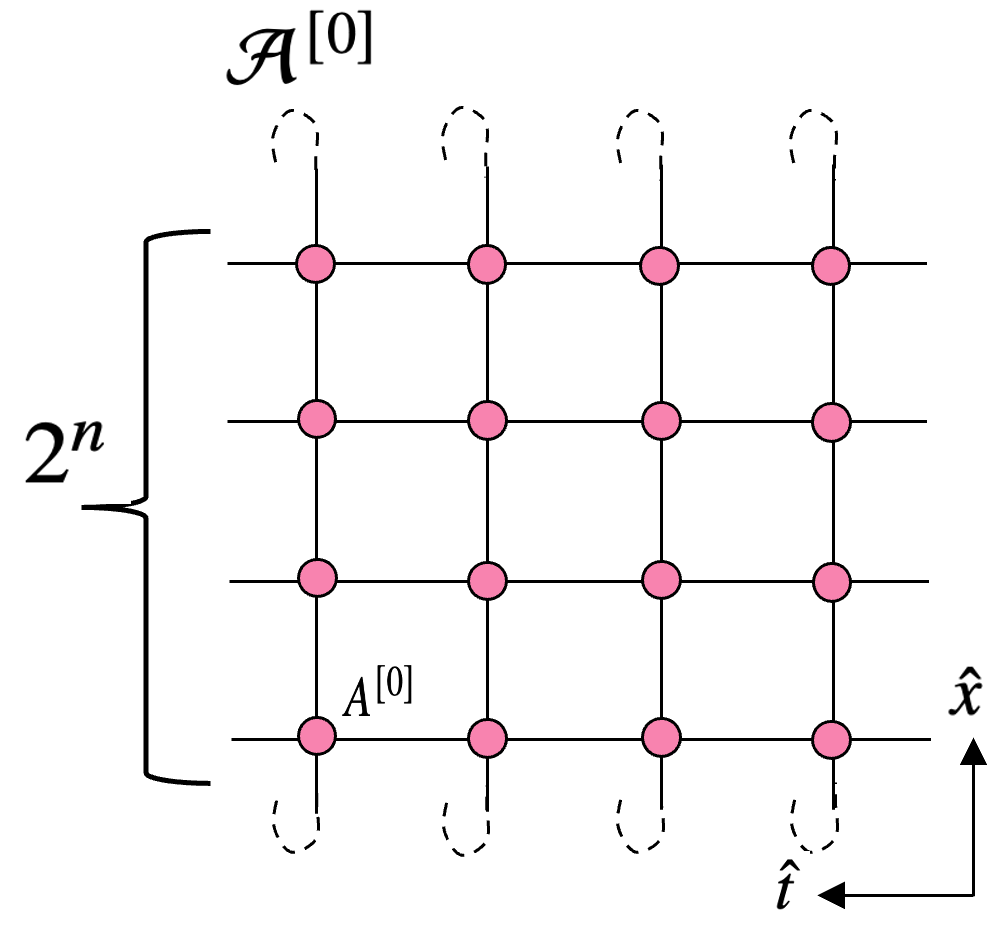}
\caption{}
\label{subfig:pure}
\end{subfigure}
\begin{subfigure}[b]{0.4\textwidth}
\centering
\includegraphics[width=4cm,height=4cm]{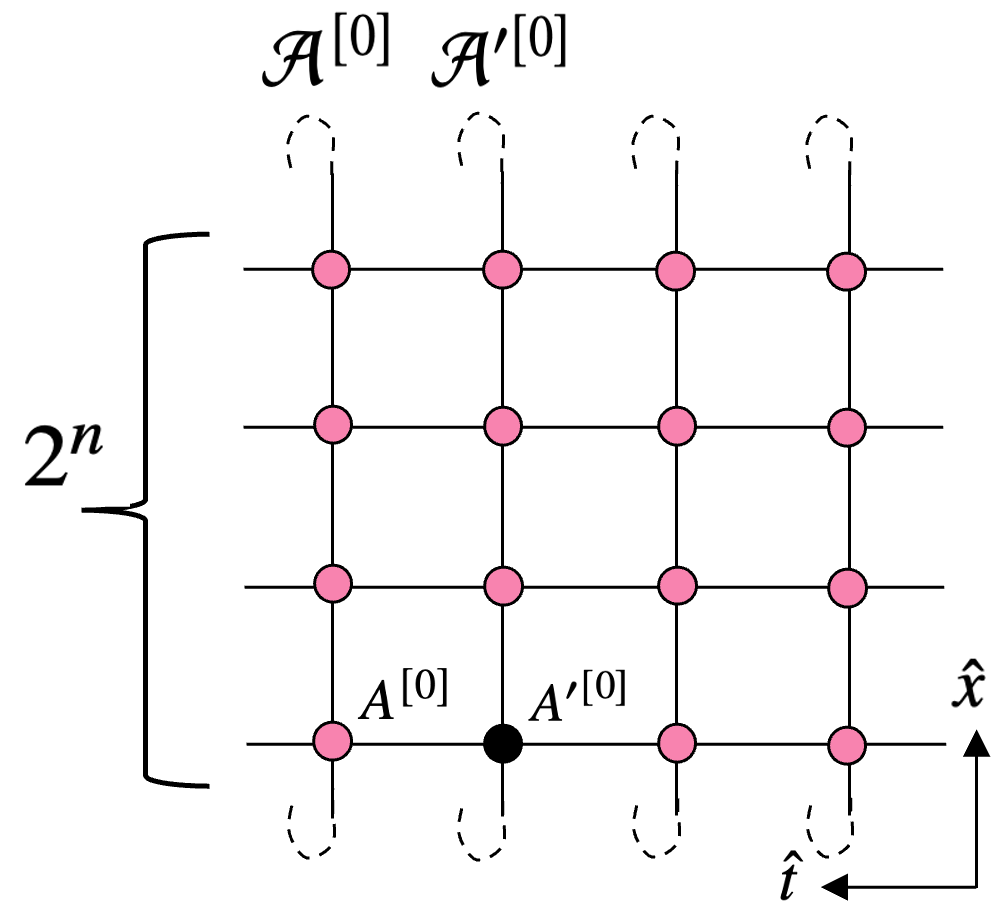}
\caption{}
\label{subfig:impure}
\end{subfigure}
\caption{(a) Initial pure tensor network.
(b) Initial impurity tensor network.}
\label{fig:coarse_graining}
\end{figure}

In order to compute the estimation of the eigenvalues $\lambda_a$, 
we coarse grain a square tensor network with volume $V=2^n\times2^n$ 
 using HOTRG \cite{PhysRevB.86.045139} as shown by Fig.~\ref{subfig:pure}.
We denote the initial tensor as $A^{[0]}$. 
After $(n-1)$ times coarse graining,
the system consist of tensor ${A}^{[n-1]}$ with volume is reduced to $V=2\times2$.
Furthermore, we compute $\mathcal{A}^{[n]}$
by applying direct contraction to tensors $A^{[n-1]}$ as follows: 
\begin{equation}
\mathcal{A}^{[n]}_{k_1k_2,j_1j_2}=\sum_{l_1,l_2,m_1,m_2,n_1,n_2}
A_{k_1l_1n_1l_2}^{[n-1]}A^{[n-1]}_{k_2l_2n_2l_1}
A_{n_1m_1j_1m_2}^{[n-1]}A_{n_1m_2j_1m_1}^{[n-1]}.
\end{equation}
The dimension of $\mathcal{A}^{[n]}$ is $(\chi^2,\chi^2)$ and its diagonalization is given by
\begin{equation}
\mathcal{A}^{[n]}=W^{[n]}\lambda^{[n]}W^{[n]\dagger}
\end{equation} 
where $W^{[n]}$ is a unitary matrix consisted of approximate eigenvectors
and $\lambda^{[n]}$ is the estimation of the eigenvalues of transfer matrix 
$\lambda_a\approx \left(\lambda_a^{[n]}\right)^{1/L_t}$.
The factor $1/L_t$ appears because 
we coarse grain the two-dimensional tensor network $\left(\mathcal{A}^{[0]}\right)^{L_t}$
instead of one time slice tensor network $\mathcal{A}^{[0]}$.
Lastly, using $\lambda^{[n]}$, the numerical estimate of energy gap can be written as
\begin{equation}\label{eq_energy_gap}
\omega_a\approx\frac{1}{L_t}\log \left(\frac{\lambda_0^{[n]}}{\lambda_a^{[n]}}\right)
=\omega_a^{[\text{hotrg}]}, 
~~~~\text{for $a=0,1,\ldots,\chi^2-1$.}
\end{equation}
\subsection{Identification of quantum number and momentum}
In this section we will show how to identify the quantum number $q_a$ of the state $|a\rangle$ 
using the matrix elements of a proper operator $\hat{\mathcal{O}}_q$ defined as
\begin{equation}\label{eq_Bba}
B_{ba}\coloneqq \langle b|\hat{\mathcal{O}}_q|a\rangle=\left(U^{\dagger}{\mathcal{O}}_qU\right)_{ba}.
\end{equation}
The selection rule associated with the symmetry of the system 
is needed to identify the quantum number.
The symmetry of (1+1)d Ising model is $Z_2$ symmetry 
so that there are two possible quantum number i.e. $q=\pm 1$.
If $\hat{D}$ is a discrete transformation operator that commutes with Hamiltonian and
has property $\hat{D}\hat{D}^{-1}=1$, then the matrix elements becomes
\begin{equation}\label{eq_selectionrule}
B_{ba}=\langle b|\hat{D}\hat{D}^{-1}\hat{\mathcal{O}}_q\hat{D}\hat{D}^{-1}|a\rangle
=q_bq_aq\langle b|\hat{\mathcal{O}}_q|a\rangle,
\end{equation}
where $q_a$ is from $\hat{D}|a\rangle=q_a|a\rangle$ and $q$
is from $\hat{D}^{-1}\hat{\mathcal{O}}_q\hat{D}=q\hat{\mathcal{O}}_q$.
From eq.~(\ref{eq_selectionrule}), we have selection rule:
\begin{equation}\label{eq_select_rule}
\text{if\space}B_{ba}=\langle b|\hat{\mathcal{O}}_q|a\rangle \neq 0 \Rightarrow q_bq_aq=1.
\end{equation}

This matrix elements $B_{ba}$ has the same dimension as the transfer matrix.
The direct computation of $B_{ba}$ is computationally expensive so that
in a similar way with the transfer matrix,
we firstly present $B_{ba}$ in terms of tensor network: 
\begin{equation}\label{eq_Bba_TN}
B_{ba}=\left(\lambda^{-m+1/2}W^{\dagger}\mathcal{A}^{m-1}
\mathcal{A}'\mathcal{A}^{m}W\lambda^{-m-1/2}\right)_{ba}.
\end{equation}
The exponent $m$ in eq.~(\ref{eq_Bba_TN}) is choosen to be  $m=\frac{L_t}{2}$.
The estimation of $\lambda$ and $W$ in eq.~(\ref{eq_Bba_TN}) can be obtained 
from coarse graining the pure tensor network in Fig.~\ref{subfig:pure}.
Meanwhile, the estimation of $\mathcal{A}^{m-1}\mathcal{A}'\mathcal{A}^{m}$ is obtained 
through coarse graining the impurity tensor network in Fig.~\ref{subfig:impure}.
Note that $\mathcal{A}'$ is one time slice tensor network
containing impurity tensor $A'$.
For a single spin operator $\hat{\mathcal{O}_q}=s_x$,
impurity tensor $A'$ is defined as
 $A'_{k_xl_xj_xl_{x-1}}\coloneq \sqrt{\sigma_{k_x}\sigma_{l_x}\sigma_{j_x}\sigma_{l_{x-1}}}
\sum_{s_x}s_x(u^{\dagger})_{k_xs_x}(u^{\dagger})_{l_xs_x}
u^{\dagger}_{s_xj_x}u^{\dagger}_{s_xl_{x-1}}$.
After coarse graining, the matrix elements can be expressed as follows:
\begin{equation}\label{eq_matrix_elements_hotrg}
B_{ba}\approx\left(\left(\lambda^{[n]}\right)^{-m+1/2}W^{[n]\dagger}
\mathcal{A}'^{[n]}W^{[n]}\left(\lambda^{[n]}\right)^{-m-1/2}\right)_{ba}\coloneq B_{ba}^{[\text{hotrg}]}.
\end{equation}

This technique can also be used to identify the momentum of the energy eigenstate
using momentum operator $\hat{\mathcal{O}}(p)$. For a given momentum $p$, 
if $\langle \Omega|\hat{\mathcal{O}}(p)|a\rangle\neq 0$, 
where $|\Omega\rangle$ is vacuum,
then $p$ is judged to be the momentum of $|a\rangle$.
\section{Numerical Result}
\begin{figure}[t!]
\centering
\begin{subfigure}[b]{0.3\textwidth}
\centering
\includegraphics[width=5.6cm,height=4.5cm]{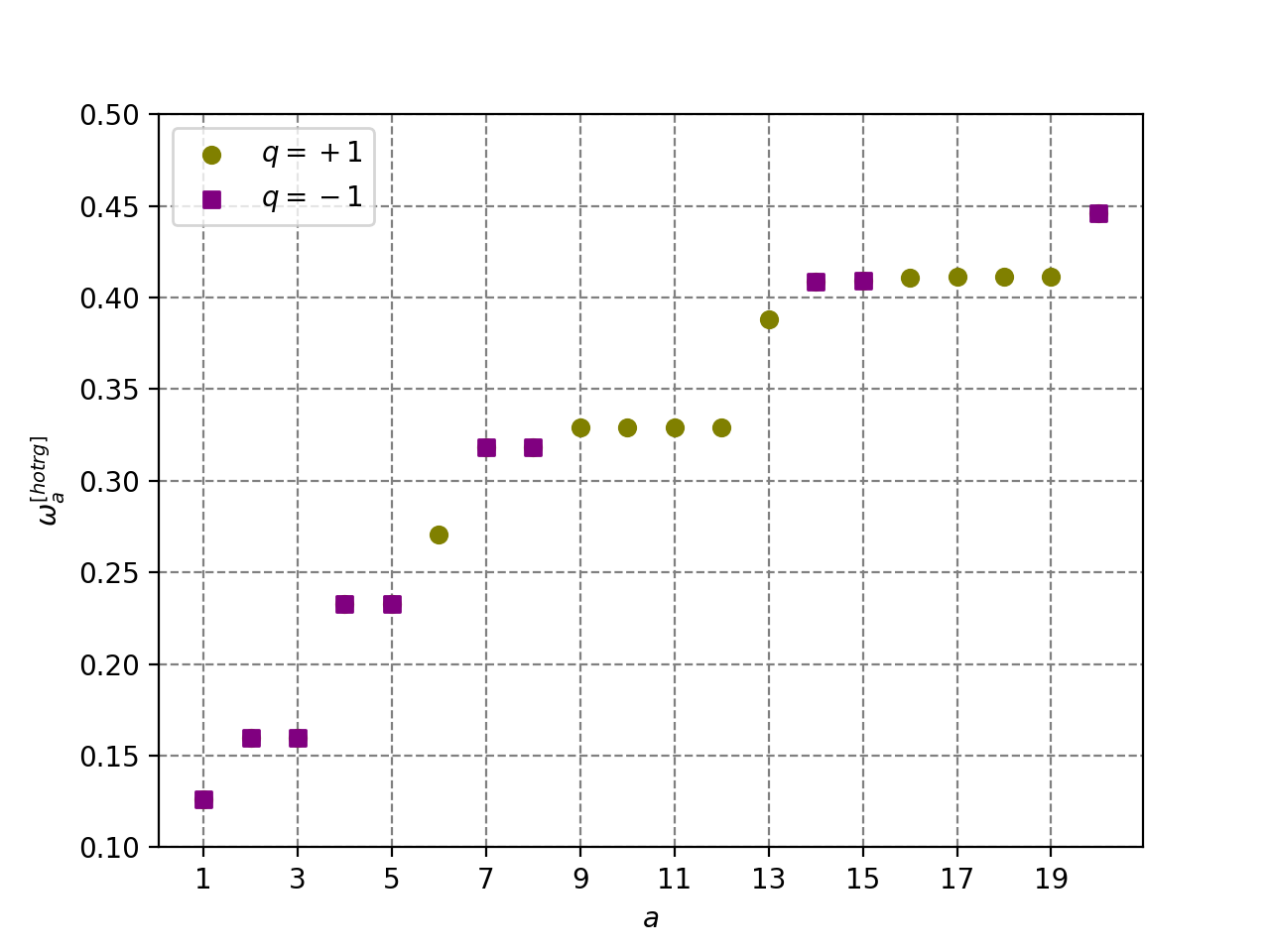}
\caption{}
\label{subfig:energy_spectrum}
\end{subfigure}\hspace{6mm}
\begin{subfigure}[b]{0.3\textwidth}
\centering
\includegraphics[width=5.6cm,height=4.5cm]{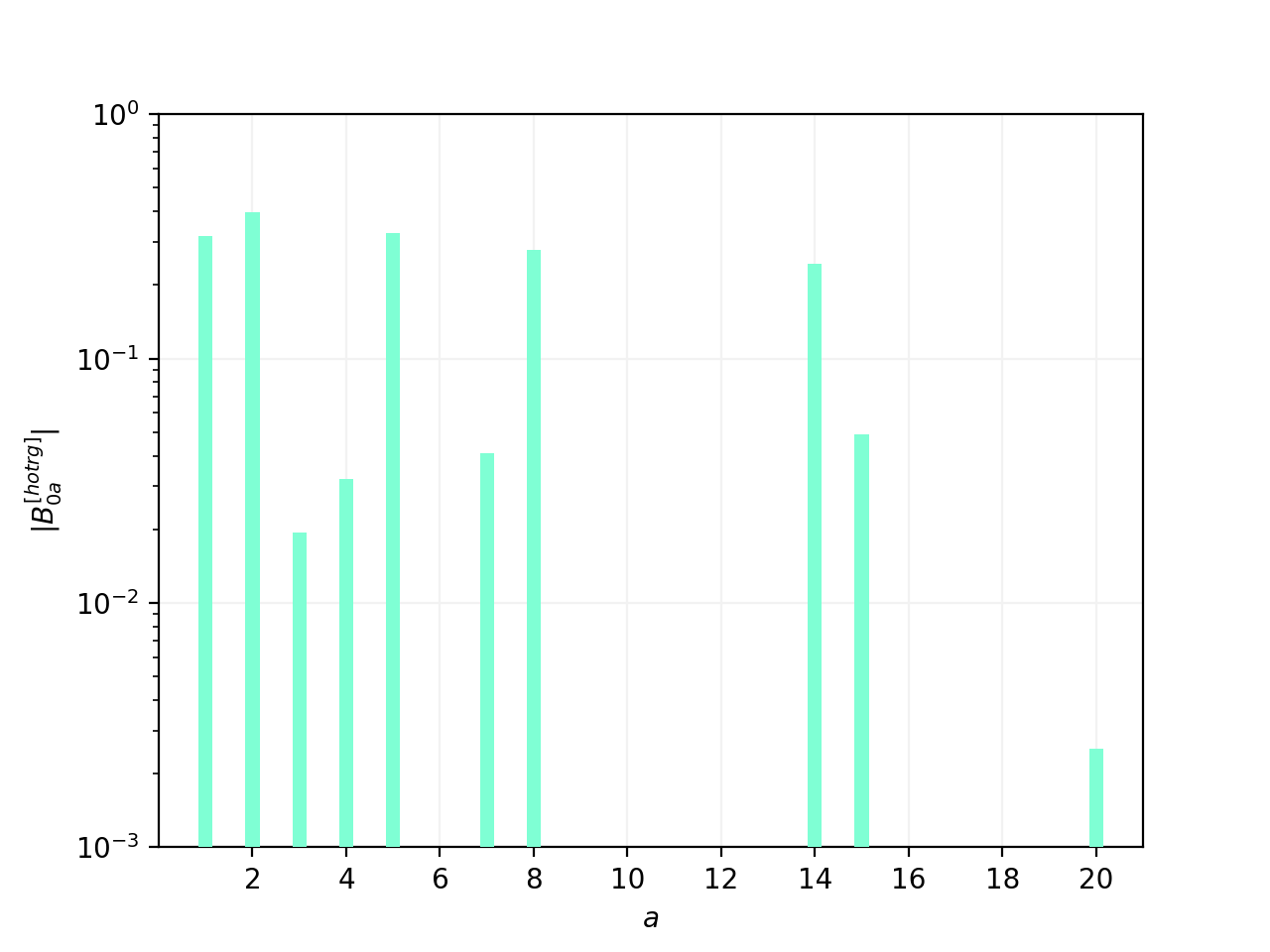}
\caption{}
\label{subfig:matrix_elements}
\end{subfigure}\hspace{6mm}
\begin{subfigure}[b]{0.3\textwidth}
\centering
\includegraphics[width=5.6cm,height=4.5cm]{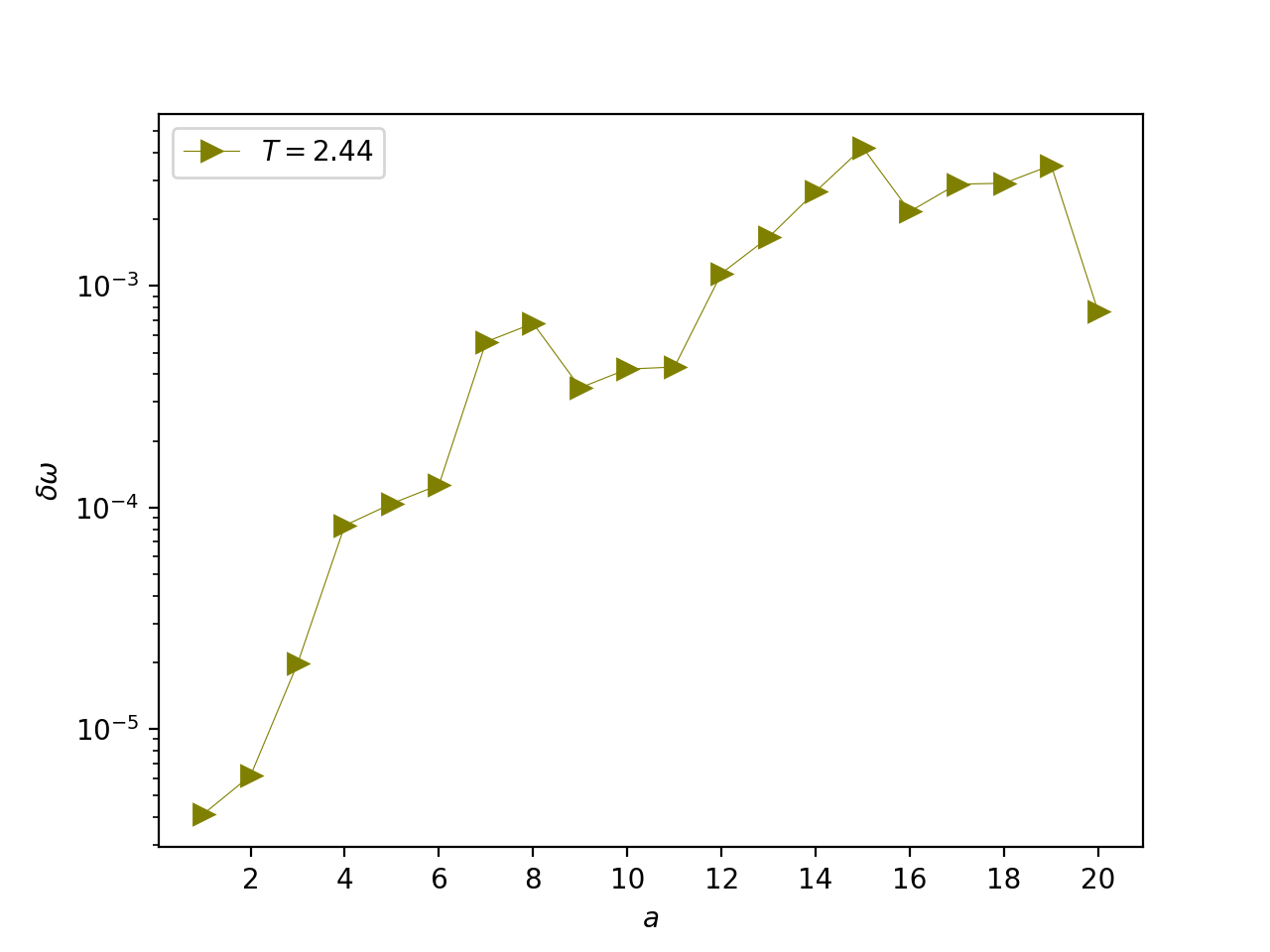}
\caption{}
\label{subfig:error_energy}
\end{subfigure}
\caption{The numerical result of energy gaps and matrix elements of system at $T=2.44$ 
with size $L_x=64$ computed by HOTRG with $\chi=80$.
(a) The energy gaps $\omega_a^{[\rm hotrg]}$ over eigenstates $a=1,2,\ldots,20$. 
(b) The matrix elements $B_{0a}^{[\rm hotrg]}\approx \langle \Omega|s_0|a\rangle$. 
(c) The relative error of the energy gaps $\omega_a^{[\rm hotrg]}$.  }
\label{fig:energy_spectrum}
\end{figure}

For the demonstration of this method, 
we carry out the spectroscopy calculation of (1+1)d Ising model at higher temperature side i.e. $T=2.44$
for system size $L_x=64$ with the cut-off bond dimension $\chi=80$. 
We compute the energy gaps $\omega^{[\text{hotrg}]}_a$ 
and the result  for $a=1,\ldots,20$ is shown in Fig.~\ref{subfig:energy_spectrum}.
From this figure, we see the energy eigenstates are divided into two groups of quantum number $q=\{+1,-1\}$.
The classification is based on matrix elements $B_{0a}^{[\text{hotrg}]}$ in Fig.~\ref{subfig:matrix_elements}.
We use eq.~(\ref{eq_matrix_elements_hotrg}),
with the spin operator $\hat{\mathcal{O}}_q=s_0$ whose quantum number is $q=-1$
and the state $\langle b|$ is chosen to be vacuum $\langle \Omega|$ with $q_{\Omega}=+1$.
From eq.~(\ref{eq_select_rule}), it is concluded that if $ \langle \Omega|s_0|a\rangle \approx B_{0a}^{[\text{hotrg}]} \neq 0$
then the quantum number of $|a\rangle$ is $q_a=-1$.
On the other hand, if $B_{0a}^{[\text{hotrg}]}=0$, 
then we judge the quantum number of $|a\rangle$ to be $q_a=+1$.
From this analysis, we see that for temperature $T=2.44$ and system size $L_x=64$, the eigenstate  $a=1,2,3,4,5,7,8,14,15$ and $20$ have quantum number $q_a=-1$,
and the rest have $q_a=+1$.

In Fig.~\ref{subfig:error_energy}, 
we show the relative error of numerical energy gaps over exact energy gaps 
using  $\delta\omega_a=\frac{|\omega_a^{[\text{exact}]}-\omega_a^{[\text{hotrg}]}|}{\omega_a^{[\text{exact}]}}$.
We compute $\omega_a^{[\text{exact}]}$ by following formulation in \cite{PhysRev.76.1232}.
The result shows that $\delta\omega_a$ tends to be larger for higher eigenstate 
and the largest error is in order $O(10^{-2})$.

To identify the momentum, we compute the matrix elements
\begin{equation}\label{eq_mat_momentum}
\langle \Omega|\hat{\mathcal{O}}_1(p)|a\rangle=
\langle\Omega|\frac{1}{L_x}\sum_{x=0}^{L_x-1} s_x e^{-ipx}|a\rangle
\approx B_{0a}(p)^{[\text{hotrg}]},
\end{equation}
using HOTRG with $\chi=80$.
For a finite system, the momentum is discretized into $p=\frac{2\pi n}{L_x}$ 
with $n=0,1,2,\ldots,L_x-1$. 
By carefully examining the matrix elements $B_{0a}(p)^{\text[\rm hotrg]}$, 
we conclude that the eigenstate $|a\rangle$ with $q=-1$ in Fig.~\ref{subfig:energy_spectrum}
has momentum as follows: 
$|0\rangle, |20\rangle \Rightarrow |p|=0$; 
$|2\rangle, |3\rangle \Rightarrow |p|=\frac{2\pi}{L_x}$; 
$|3\rangle ,|4\rangle \Rightarrow |p|=\frac{4\pi}{L_x}$; 
$|5\rangle, |6\rangle\Rightarrow |p|=\frac{6\pi}{L_x}$; and 
$|7\rangle,|8\rangle\Rightarrow |p|=\frac{8\pi}{L_x}$.
\begin{figure}[t!]
\centering
\begin{subfigure}[b]{0.4\textwidth}
\centering
\includegraphics[width=6cm,height=4.5cm]{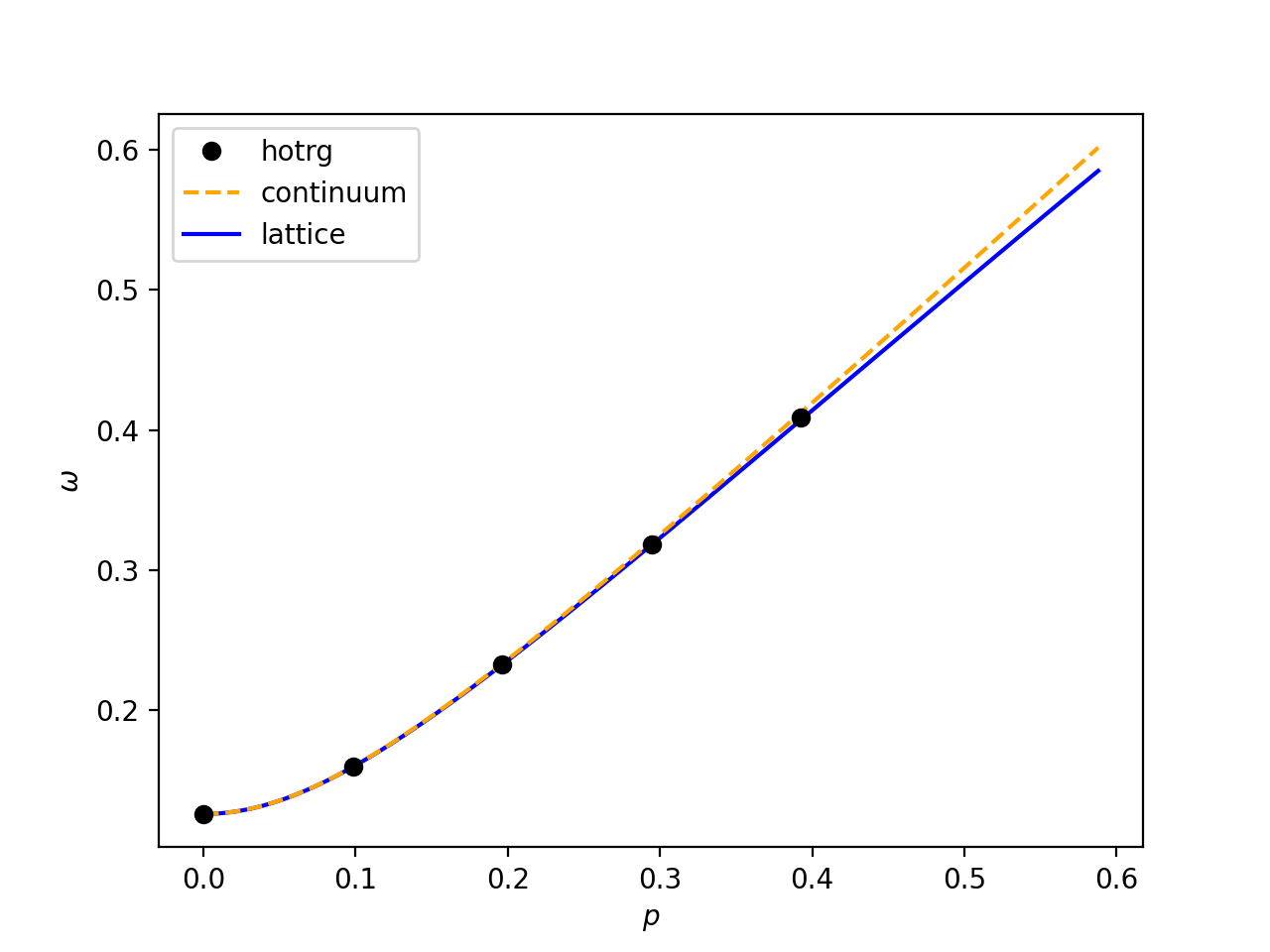}
\caption{}
\label{subfig:dispersion_relation}
\end{subfigure}\hspace{20mm}
\begin{subfigure}[b]{0.4\textwidth}
\centering
\includegraphics[width=6cm,height=4.5cm]{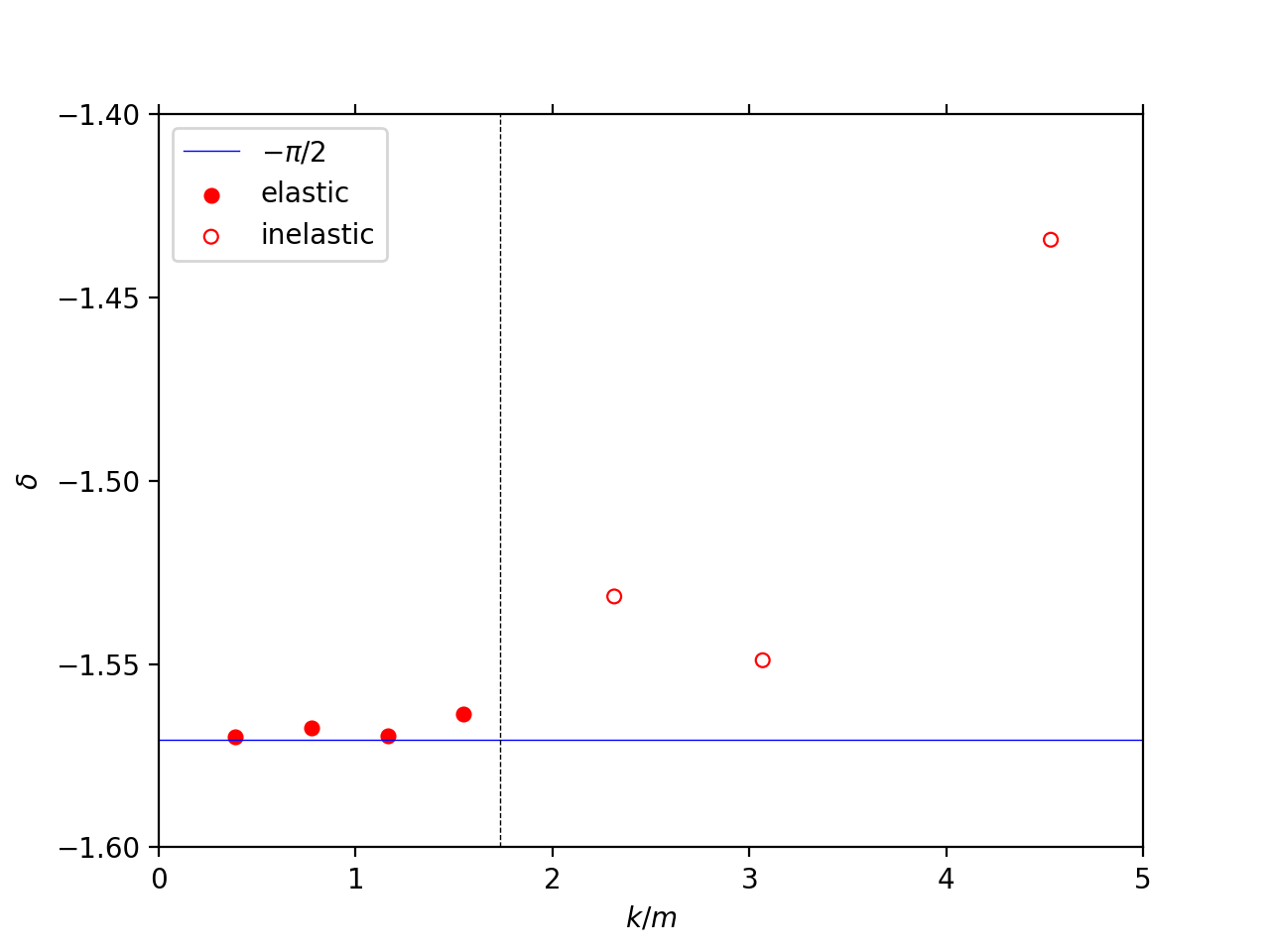}
\caption{}
\label{subfig:phaseshift}
\end{subfigure}
\caption{(a) The dispersion relation of energy and momentum.
(b) The numerical scattering phase shift of $(q+1)$d Ising model obtained using our method.}
\end{figure}
The energy $\omega_a^{[\rm hotrg]}$ and momentum $p$ is plotted
into dispersion relation in Fig.~\ref{subfig:dispersion_relation}.
The figure shows that the dispersion relation from HOTRG data agree with the continuum $\omega=\sqrt{m^2+p^2}$
and the lattice dispersion relation $\omega=\arccosh(1-\cos p + \cosh m)$ \cite{Gattringer:1992np}.

Furthermore, we also identify the two particle states with total momentum $P=0$ 
using the matrix elements of the double-spin operator in the momentum space as follows:
\begin{equation}\label{eq_matrix_elements_2particle}
\langle \Omega|\hat{\mathcal{O}}_2(P,p)|a\rangle
=\langle\Omega|\frac{1}{L_x^2}\sum_{x,y=0}^{L_x-1} s_x s_ye^{-ip_1x}e^{-ip_2y}|a\rangle
\approx B_{0a}(P,p)^{[\text{hotrg}]},
\end{equation}
where $p_j=\frac{2\pi n_j}{L_x}$ for $j=1,2$ and $n_j=0,1,2,\ldots, L_x-1$ 
are momentum of the first and second particle respectively,
$P=p_1+p_2$ is the total momentum, and
$p=(p_1-p_2)/2$ is the relative momentum.
If $\langle \Omega|\hat{\mathcal{O}}_2(P,p)|a\rangle\neq 0$, 
then $|a\rangle$ is assigned to be the two-particle state 
with total momentum $P$ irrespective of $p$.
In this research, we only consider the two-particle state whose total momentum is $P=0$.
We compute $B_{0a}^{[\text{hotrg}]}(P,p)$ at $T=2.44$ 
for several system size $L_x=8-64$ with $\chi=80$,
and the results are shown in Table \ref{tab:total_zero_momentum_even_sector}.
\begin{table}[t]
\begin{center}
\caption{The energy gaps $\omega_a^{[\text{hotrg}]}$ of two-particle state at system size $L_x=8-64$,
and their matrix elements $B_{0a}(P,p)^{[\text{hotrg}]}$ at total momentum $P=0$ and $P\neq 0$.}
\label{tab:total_zero_momentum_even_sector}
\begin{tabular}{rrrrrr}
\hline
$L_x$ & $a$ &$\omega_a^{[{\rm hotrg}]}$  & $B_{0a}(0,0)^{[\text{hotrg}]}$ 
& $B_{0a}(0,\frac{2\pi}{L_x})^{[\text{hotrg}]}$& $B_{0a}(\frac{2\pi}{L_x},\frac{\pi}{L_x})^{[\text{hotrg}]}$\\
\hline\hline
 8&  4& $0.814585$ & $0.37740$ & $0.12364$ & $<10^{-15}$ \\
  & 19& $2.133922$ & $0.07730$ & $0.04844$ & $<10^{-12}$ \\
\hline
16&  4& $0.465348$ & $0.31004$ & $0.09529$ & $<10^{-15}$ \\
  & 18& $1.171480$ & $0.06904$ & $0.05901$ & $<10^{-12}$ \\
\hline
32&  4& $0.319553$ & $0.21122$ & $0.06541$ & $<10^{-14}$ \\
  & 14& $0.636356$ & $0.04705$ & $0.06178$ & $<10^{-10}$ \\
\hline
64&  6& $0.270836$ & $0.12007$ & $0.03888$ & $<10^{-14}$ \\
  & 13& $0.387849$ & $0.03007$ & $0.05024$ & $<10^{-9}$ \\
\hline
\end{tabular}
\end{center}
\end{table}
The energy $\omega_a^{[{\rm hotrg}]}$ 
listed in Table \ref{tab:total_zero_momentum_even_sector} are judged 
to be the energy of two-particle state with the total momentum is equal to zero
because the matrix elements $B_{0a}(P,p)^{[\text{hotrg}]}$ of the corresponding state
in each system size $L_x$ are non-zero when $P=0$ 
while the values become extremely small or considered as zero when $P\neq 0$.
Using the energy in Table \ref{tab:total_zero_momentum_even_sector},
we compute the relative momentum $k$,
\begin{equation}
\omega=2\sqrt{k^2+m^2}
\end{equation}
where $m$ is the rest mass 
and here we use the exact one at infinite volume limit $m=0.12621870$.
Using this $k$, the phase shift $\delta(k)$ is computed from L\"uscher's formula \cite{Luscher:1986pf},
\begin{equation}
e^{2i\delta(k)}=e^{-ikL}.
\end{equation}
The plot of $\delta(k)$ over $k/m$ is given in Fig.~\ref{subfig:phaseshift},
where we divide the plot into the elastic and the inelastic area.
The dashed line represents the limit of the elastic region 
which is defined as $0 \leq k/m <\sqrt{3}$. 
The phase shift in the elastic region is around $-\pi/2$
which agrees with theoretical prediction 
i.e. $\delta_{\text{ising}}=-\pi/2$ \cite{Gattringer:1992np}.

\section{Summary and outlook}
We propose a spectroscopy scheme using the transfer matrix and the tensor network,
and demonstrate it with $(1+1)$d Ising model.
The computation is done at temperature $T=2.44$ 
and system size $L_x=64$ using HOTRG with $\chi=80$.
In this parameter set, 
our method can reproduce the energy gaps $\omega_a^{[\text{hotrg}]}$ and its quantum number
correctly up to $a=20$ with the largest relative error is in order $O(10^{-2})$.
The low-lying momentum of one-particle channel is identified 
by computing the matrix elements of the single spin operator in the momentum space.
Furthermore, two particle states with the total momentum $P=0$ 
at several system sizes $L_x=8,16,32,64$ are identified 
by examining the matrix elements of double-spin operator in the momentum space.
Using the energy gaps of two-particle states and the L\"uscher's formula,
the scattering phase shift is determined.
The numerical phase shift in elastic region agrees with theoretical result i.e. $\delta=-\pi/2$.
For future work, we will apply this method to (1+1)d scalar field theory,
and also use the method to compute the phase shift in the moving frame.
\acknowledgments
A. I. F. is supported by MEXT, JICA, and JST SPRING, Grant No. JPMJSP2135.
S.T. is supported in part by JSPS KAKENHI Grants No.~21K03531, and No.~22H01222.
T.Y. is supported in part by JSPS KAKENHI Grant No.~23H01195 and MEXT as ``Program for Promoting Researchers on the Supercomputer Fugaku'' (Grant No. JPMXP1020230409).
This work was supported by MEXT KAKENHI Grant-in-Aid for Transformative Research Areas A ``Extreme Universe'' No. 22H05251.

\end{document}